\documentclass{pnastwo}
\usepackage{pnastwoF}
\usepackage{amssymb,amsfonts,amsmath}
\def\p100{P_{<100}}
\begin{document}
\title{The estimate of the wind energy potential and insolation}
\author{Kenichiro Aoki\affil{1}{Research and Education Center for
    Natural Sciences and Hiyoshi Dept. of Physics, Keio University,
    Yokohama 223--8521, Japan }}
\contributor{Submitted to Proceedings of the National Academy of Sciences
of the United States of America}
\maketitle
\begin{article}    
In a solid and detailed paper, Lu, McElroy and Kiviluoma computed the
potential for wind-generated electricity\cite{Lu}.  We believe such
estimates are both interesting and important. However, the estimate
seems too high, essentially because the wind turbines have direct
access to only a small vertical fraction of the atmosphere.

The average amount of solar radiation at the top of the atmosphere is
340\,W/m$^2$.  Assuming that 1\,\% of this is in the kinetic energy of
the atmosphere\cite{Lu} and since 1\,\% of the atmospheric mass is in
the bottom 100\,m layer of air, we arrive at $\p100=$0.034\,W/m$^2$,
which is roughly the {\it total kinetic energy of the atmosphere} per
area in this layer.  We chose the layer relevant to the 100\,m
diameter turbines at low altitudes considered in the paper.  Compare
this to the wind farm numbers used in the work; 2.5\,MW turbines with
an interturbine aerial spacing of 0.28\,km$^2$ (onshore)\cite{Lu}
corresponds to 1.8\,W/m$^2$ at 20\,\% capacity factor, which is the
minimum capacity factor used in the work.  This is 53 $\times \p100$.
Let us also look at the final result in Table 1 in \cite{Lu}: Taking
the example of Germany (onshore, area:$3.6\times10^{11}$\,m$^2$), the
overall average for the wind energy potential corresponds to
1.0\,W/m$^2 = 29\times\p100$. This is lower than the above value and
consistent with it, since the authors were carefully selective of the
regions for deployment, with due reason.

The estimates of \cite{Lu} are higher than $\p100$ by much more than
an order of magnitude, as explained above.  We believe that for such
an amount of energy to be extracted, large back reaction to change the
air circulation would at least need to occur, which is a crucial
aspect not considered there.  Our calculations are rough: For
instance, we did not subtract for albedo nor account for the higher
wind velocities at higher altitudes taking a larger fraction of the
atmospheric kinetic energy, so that the discrepancy could quite well
be larger.  We believe it is important to understand the mechanism
behind such a large discrepancy to clarify whether it is possible to
extract such an amount of energy from wind.

\end{article}

\end{document}